\documentclass[conference]{IEEEtran}
\usepackage{amssymb}
\usepackage{graphicx,subfigure}
\usepackage[centertags]{amsmath}
\newtheorem{theorem}{Theorem}
\newtheorem{lemma}{Lemma}

\long\def\comment#1{}
\date{}
\begin{document}
\renewcommand{\textfraction}{0}
\title{The Single Source Two Terminal Network with Network Coding}

\author{\authorblockN{Aditya Ramamoorthy}
\authorblockA{Department of Electrical Engineering\\
University of California\\
Los Angeles, CA 90095\\
Email: adityar@ee.ucla.edu} \and
\authorblockN{Richard D. Wesel}
\authorblockA{Department of Electrical Engineering\\
University of California\\
Los Angeles, CA 90095\\
Email: wesel@ee.ucla.edu}}

\maketitle \thispagestyle{empty}
\begin{abstract}
We consider a communication network with a single source that has
a set of messages and two terminals where each terminal is
interested in an arbitrary subset of messages at the source. A
tight capacity region for this problem is demonstrated. We show by
a simple graph-theoretic procedure that any such problem can be
solved by performing network coding on the subset of messages that
are requested by both the terminals and that routing is sufficient
for transferring the remaining messages.
\end{abstract}
\normalsize
\section{Introduction}
The seminal work of Ahlswede {\it et al.} \cite{al} established
that for the single-source multiple-terminal multicast problem the
achievable rate was the minimum of the maximum flows to each
terminal from the source. They showed that in general, it is
necessary to perform network coding to achieve this capacity. The
basic idea is to give the nodes in the network the flexibility of
performing operations on the data rather than simply replicating
and/or forwarding it. Li {\it et al.} \cite{lc} showed that linear
network coding is sufficient for achieving the capacity of the
transmission of a single source to multiple terminals. Subsequent
work by Koetter and M\'{e}dard \cite{rm} and Jaggi {\it et al.}
\cite{jaggipoly} presented constructions of linear multicast
network codes. A randomized construction of multicast codes was
demonstrated by Ho {\it et al.} \cite{tracey_rand}.

It is important to realize that the multicast capacity result of
\cite{al} assumes that all the terminals are interested in the
same data. The general network coding problem with multiple
sources and terminals and an arbitrary set of connections is much
harder and not much is known about it. In fact it has been shown
in \cite{dougherty} that non-linear network codes are necessary in
certain non-multicast problems. Network coding has also been
considered from a lossless compression point of view in
\cite{effrosdimacs}\cite{tracey_ciss}\cite{adimsr}\cite{adimsrjnl}.

In this paper we study a specific example of a non-multicast
problem with a single source and two sinks. We find a tight
capacity region for this problem. This problem was independently
considered by Ngai and Yeung \cite{ngaiyeung} and Erez and Feder
\cite{erez}\footnote{We became aware of this work after the
submission of the current paper.}. However our method of proof is
very different and is based on a simple graph-theoretic procedure
that may be of independent interest. This procedure was also
utilized in \cite{adimsrjnl}.

\section{Problem Formulation}
Consider a communication network modelled as a directed graph $G$,
with a specified source node $S$ and two terminal nodes $T_1$ and
$T_2$. We assume that the links are noiseless and that each edge
in $G$ has unit capacity. This assumption can be realized by
picking a suitably large time unit, assuming sufficient
error-correction at the lower layers of the network and splitting
edges of higher capacity into parallel unit capacity edges.

Suppose that the source node $S$ observes three independent
processes $X_0, X_1$ and $X_2$ such that terminal $T_1$ is
interested in $(X_0, X_1)$ and terminal $T_2$ is interested in
$(X_0, X_2)$. Let the entropy rates of the processes be $H_0, H_1$
and $H_2$ respectively. We show the necessary and sufficient
conditions for the feasibility of this connection. Furthermore it
is shown that this problem can be solved by a combination of pure
routing and network coding, where the sources $X_1$ and $X_2$ can
be simply routed to $T_1$ and $T_2$ whereas the source $X_0$ may
need network coding. The case of connections between terminal
nodes is handled more naturally in our framework as compared to
\cite{ngaiyeung}.

In the sequel the capacity assignment to an edge $a \rightarrow b$
is denoted by $cap~(a \rightarrow b)$ and the minimum cut between
nodes $V_1$ and $V_2$ is denoted by $min\mbox{-}cut(V_1, V_2)$. By
the max-flow min-cut theorem \cite{vanlintwilson}, the minimum cut
is also the maximum rate that can be transmitted from $V_1$ to
$V_2$. By a solution to a given problem we mean an assignment of
appropriate coding vectors to each edge so that the required
network connection can be supported.
\section{Results}
\begin{figure}[t]
\centering
\includegraphics[height=9cm, clip=true]{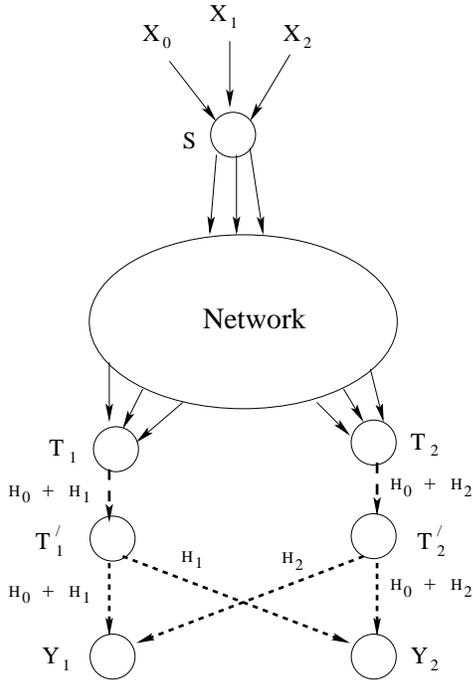}
\caption{\label{fig:cwit_aug} The figure shows the augmented graph
$G_1$. The original graph $G$ comprises of $S, T_1, T_2$ and the
network. The augmented graph $G_1$ also contains the virtual
terminals $T_1^{\prime}$ and $T_2^{\prime}$ and the nodes $Y_1$
and $Y_2$. The virtual edges are denoted by dashed lines and their
capacities are labelled. }
\end{figure}

The following theorem is the main result of this paper.
\begin{theorem}
\label{main_thm}
{\it Consider a communication network modelled by a directed graph
$G = (V, E)$ with one source node $S$ and two terminal nodes $T_1$
and $T_2$. Three independent processes $X_0, X_1$ and $X_2$ are
observed at S such that $H(X_0) = H_0, H(X_1) = H_1$ and $H(X_2) =
H_2$. $T_1$ is interested in receiving $(X_0, X_1)$ and $T_2$ is
interested in receiving $(X_0, X_2)$. If
\begin{eqnarray}
\label{ineq_1} min\mbox{-}cut (S, T_1) &\geq& H_0 + H_1, \\
min\mbox{-}cut(S, T_2) &\geq& H_0 + H_2 \textrm{~and,} \\
\label{ineq_3} min\mbox{-}cut(S, (T_1, T_2)) &\geq& H_0 + H_1 + H_2
\end{eqnarray}
there exists a solution where $X_1$ can be routed to $T_1$, $X_2$
can be routed to $T_2$ and $X_0$ can be sent to both $T_1$ and
$T_2$ via network coding. Conversely if any of the inequalities
(\ref{ineq_1}) - (\ref{ineq_3}) are violated then the connection
cannot be supported.}\\
\end{theorem}
We defer the proof of this theorem until we have established a
lemma that is required. We start by defining an augmented graph
$G_1 = (V_1, E_1)$ as depicted in Fig. \ref{fig:cwit_aug}.
\begin{enumerate}
\item The new vertex set is $V_1 = V \cup \{T_1^{\prime},
T_2^{\prime}, Y_1, Y_2\}$ as shown in Fig. \ref{fig:cwit_aug}.
$T_1^{\prime}$ and $T_2^{\prime}$ can be regarded as virtual
terminals, where the data is actually decoded. $Y_1$ and $Y_2$ are
virtual nodes introduced for the purposes of our proof.
\item The capacity assignments of the new edges are $cap~(T_1
\rightarrow T_1^{\prime}) = H_0 + H_1, cap~(T_1^{\prime}
\rightarrow Y_1) = H_0 + H_1, cap~(T_1^{\prime} \rightarrow Y_2) =
H_1, cap~(T_2 \rightarrow T_2^{\prime}) = H_0 + H_2,
cap~(T_2^{\prime} \rightarrow Y_1) = H_2$ and $cap~(T_2^{\prime}
\rightarrow Y_2) = H_0 + H_2$.
\end{enumerate}
\begin{lemma}
For the augmented graph $G_1$ the following is true :-
\begin{eqnarray}
min\mbox{-}cut(S, T_1^{\prime}) &=& H_0 + H_1 \\
\label{mincut_t_2_prime} min\mbox{-}cut(S, T_2^{\prime}) &=& H_0 + H_2 \\
min\mbox{-}cut(S, (T_1^{\prime}, T_2^{\prime})) &\geq& H_0 + H_1 + H_2 \\
\label{mincut_y_1} min\mbox{-}cut (S, Y_1) &=& H_0 + H_1 + H_2 \\
min\mbox{-}cut (S, Y_2) &=& H_0 + H_1 + H_2
\end{eqnarray}
\end{lemma}
\emph{Proof :-}
The first two equalities are obviously true. To see that
$min\mbox{-}cut (S, Y_1) = H_0 + H_1 + H_2$ note that all cuts
between $S$ and $Y_1$ can be divided into four types:
\begin{itemize}
\item[a)] The cut $(C, C^{c})$ such that $S, T_1, T_2 \in C$ and
$Y_1 \in C^c$. By inspection such a cut has capacity larger than
or equal to $H_0 + H_1 + H_2$. \item[b)] $S, T_1 \in C$ and $T_2,
Y_1 \in C^c$.\\ The $min \mbox{-}cut (S, T_2) \geq H_0 + H_2$ and
$min \mbox{-}cut (T_1, Y_1) = H_0 + H_1$ and the edges connecting
$T_1$ and $Y_1$ are independent of the edges connecting $S$ and
$Y_1$. This means that such a cut has capacity at least $2H_0 +
H_1 + H_2$.
\item[c)] $S, T_2 \in C$ and $T_1, Y_1 \in C^c$.\\
The $min\mbox{-}cut (S, T_1) \geq H_0 + H_1$ and
$min\mbox{-}cut(T_2, Y_1) = H_2$ and the edges connecting $S$ to
$T_1$ are independent of the edges connecting $T_2$ to $Y_1$. This
means that such a cut has capacity at least $H_0 + H_1 + H_2$.
\item[d)] $S \in C$ and $T_1, T_2, Y_1 \in C^c$.\\ Since the $min
\mbox{-}cut (S, (T_1, T_2)) \geq H_0 + H_1+ H_2$, therefore any
such cut has capacity at least $H_0 + H_1+ H_2$.
\end{itemize}
Finally, the sum of the capacities on the incoming edges of $Y_1$
is exactly $H_0 + H_1 + H_2$. This means that $min \mbox{-}cut (S,
Y_1) = H_0 + H_1 + H_2$. The other statements in the lemma can be
shown to be true in a similar manner.
\endproof
\vspace{3mm}
Using the augmented graph $G_1$ we shall now demonstrate the
existence of a certain number of paths from $S$ to $T_1^{\prime}$
and $S$ to $T_2^{\prime}$ over which data can be routed. Further,
we shall show that it is possible to send the remaining data via
network coding such that the demands of each sink are satisfied.
The arguments proceed by utilizing the minimum cut conditions and
performing a simple graph-theoretic procedure on the chosen paths
in $G_1$. The details are given below.\\\\
\noindent \textbf{Proof of Theorem} \ref{main_thm} :-\\
First let us consider the paths from $S$ to $Y_1$ and $S$ to $T_2^{\prime}$.
Using Menger's theorem (see the book by van Lint \& Wilson \cite{vanlintwilson}) we can conclude that :
\begin{itemize}
\item There exists a set of $(H_0 + H_1 + H_2)$ edge-disjoint
paths from $S$ to $Y_1$ from (\ref{mincut_y_1}). We call this set
$\mathcal{G}$.
\item There exists a set of $(H_0 + H_2)$ edge-disjoint paths from
$S$ to $T_2^{\prime}$ from (\ref{mincut_t_2_prime}). We call this
set $\mathcal{R}$.
\end{itemize}
Now, we color the edges in paths $\in \mathcal{G}$, {\it green}
and the edges in paths $\in \mathcal{R}$, {\it red}. At the end of
this procedure some edges on these paths may have just one color
while others may have two.
\par We claim that it is always possible to find $H_1$ exclusively
{\it green} paths (i.e. paths that contain edges only having the
color {\it green}) from $S$ to $T_1^{\prime}$. The technique of
proof is similar to the one used in
\cite{adimsrjnl}\cite{jainsteiner}. To prove this we define an
algorithm $A$ that shall be applied to a path $P \in
\mathcal{G}$.\\\\
\textbf{Algorithm A (P)} :-
\begin{enumerate}
\item Traverse $P$ starting at $S$ and find the first edge $e_1$ that has color {\it (green, red)}
\item If no such $e_1$ is found then \textbf{STOP}.
\item \textbf{ELSE}\\
Suppose $e_1 \in P^{\prime}$ where $P^{\prime} \in \mathcal{R}$
such that $P^{\prime} = P_1^{\prime} - e_1 - P_2^{\prime}$ where
$P_1^{\prime}$ is the portion of $P^{\prime}$ from $S$ to $e_1$
and $P_2^{\prime}$ is the portion of $P^{\prime}$ from $e_1$ to
$T_2^{\prime}$. Color all edges on $P$ from $S$ to $e_1$, {\it
red} in addition to their current color and remove {\it red} from
the edges in $P_1^{\prime}$.
We now define a condition that each path $P \in \mathcal{G}$ needs to satisfy.
\begin{equation}
\begin{split}
Cond(P) &= \textrm{\{All edges in $P$ are {\it green}\}} \\
&\textrm{or \{First edge of $P$ is {\it (green, red)}\}} \\
\end{split}
\end{equation}
We continue applying $A$ to each path of $\mathcal{G}$ until all
paths in $\mathcal{G}$ satisfy $Cond$. It is easy to see that $A$
will eventually halt (for a proof see \cite{adimsrjnl}).

\par At the end of this process we realize that there exist $H_1$
paths belonging to $\mathcal{G}$ that are exclusively {\it green}.
This is true since if Algorithm $A$ re-routes a path $\in
\mathcal{R}$ it removes the color {\it red} from one outgoing edge
of $S$ and places it on another outgoing edge. Therefore the total
number of outgoing edges that are colored {\it red} remains
constant at $H_0 + H_2$. It follows that $H_0 + H_1 + H_2 - (H_0 +
H_2) = H_1$ outgoing edges are colored {\it green} and since the
paths obey $Cond$ all those paths are exclusively {\it green}.

\par Next we note that all the exclusively {\it green} paths need
to pass through $T_1^{\prime}$ since $T_2^{\prime}$ has exactly
$(H_0 + H_2)$ incoming edges all of which have to be colored {\it
red}. This proves the claim made above.
\end{enumerate}
\par The critical point to be realized is that the re-routing of
paths as above gives us $H_1$ paths from $S$ to $T_1^{\prime}$
that are interference-free since these paths do not intersect with
the paths from $S$ to $T_2^{\prime}$. This means that data on
these paths can be simply routed. Applying exactly the same
procedure on the set of paths from $S$ to $Y_2$ and $S$ to
$T_1^{\prime}$ gives us $H_2$ paths from $S$ to $T_2^{\prime}$
that are interference-free.

Now suppose that these paths ($H_1$ paths from $S$ to $T_1$ and
$H_2$ paths from $S$ to $T_2$) are removed from $G_1$ to obtain a
new graph $G_2$. Note that there still exist $H_0$ paths from $S$
to $T_1^{\prime}$ and $H_0$ paths from $S$ to $T_2^{\prime}$ in
$G_2$. In other words, even after the removal of the
interference-free paths the maximum flow from $S$ to
$T_1^{\prime}$ and $S$ to $T_2^{\prime}$ in $G_2$ is $H_0$. Using
the multicast result of \cite{al} we can surely transmit the {\it
same} $H_0$ bits from $S$ to $T_1^{\prime}$ and $T_2^{\prime}$ via
network coding. \par Thus, the entire solution can be realized by
an appropriate choice of paths such that,
\begin{enumerate}
\item $H_1$ bits (process $X_1$) can be routed from $S$ to
$T_1^{\prime}$ and $H_2$ bits (process $X_2$) can be routed from
$S$ to $T_2^{\prime}$.
\item $H_0$ bits (process $X_0$) can be
sent to both $T_1^{\prime}$ and $T_2^{\prime}$ by linear network
coding \cite{lc}.
\end{enumerate}
Finally we note that it is trivial to realize the virtual
terminals $T_1^{\prime}$ and $T_2^{\prime}$ at the terminals.

\par
\begin{figure}[t]
\centering
\includegraphics[height=6cm, clip=true]{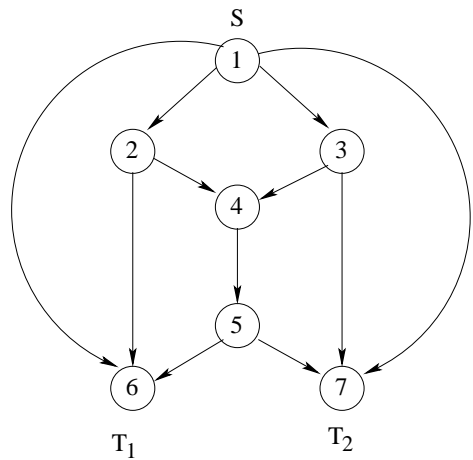}
\caption{\label{fig:tight_eg} The sources observed at $S$ are such
that $H_0 = 2, H_1 = H_2 = 1$. The figure shows a network where it
is necessary to send $X_0$ via network coding. All links have unit
capacity.}
\end{figure}

 The proof of the converse is easy to see since even if one
of the inequalities (\ref{ineq_1}) - (\ref{ineq_3}) is violated
then at least one terminal does not have enough capacity to
support its demand. This completes the proof of Theorem
\ref{main_thm}.
\endproof
It is possible to find networks where one needs to strictly
perform network coding for transmitting $X_0$ (while routing $X_1$
and $X_2$) and hence our result is tight. A simple example that
demonstrates this is provided in Fig. \ref{fig:tight_eg}. Here we
have $H_0 = 2$ and $H_1 = H_2 = 1$. In Fig. \ref{fig:tight_eg}
note that the $min\mbox{-}cut (S, (T_1, T_2)) = 4$. Therefore
among the outgoing links from $S$ namely $1 \rightarrow 6, 1
\rightarrow 2, 1 \rightarrow 3, 1 \rightarrow 7$, one link needs
to carry $X_1$, one link needs to carry $X_2$ and the remaining
two links can carry a combination of the bits from $X_0$. By the
rate requirements at the terminal it is easy to see that the
combination of the $X_0$'s needs to be carried on links $ 1
\rightarrow 2$ and $1 \rightarrow 3$. This means that the solution
needs to be realized by routing $X_1$ on link $1 \rightarrow 6$,
routing $X_2$ on link $1 \rightarrow 7$ and using the remaining
part of the network to transmit $X_0$. However the remaining part
of the network is precisely the celebrated butterfly example of
\cite{al} and we know that network coding is essential for
transmitting $X_0$ over it.

\section{Conclusion}
We found the capacity region for a network information transfer
problem with a single source and two terminals when the use of
network coding is permitted by utilizing a simple graph-theoretic
procedure that may be of independent interest. It is interesting
to note that the use of network coding permits us to obtain a
tight characterization of the capacity region of this problem.
However the region for the general broadcast channel with two
receivers is still unknown (this was also noted by \cite{erez}).
\section{Acknowledgement}
This work was supported by the University of California and Texas
Instruments through UC Discovery grant COM04-10155 and National Science
Foundation grant CCR-0209110.
\bibliographystyle{IEEEtran}

\end{document}